\documentclass[sigconf]{acmart}

\AtBeginDocument{%
  }

\copyrightyear{2026}
\acmYear{2026}
\setcopyright{cc}
\setcctype{by}
\acmConference[CAIS '26]{ACM Conference on AI and Agentic Systems}{May 26--29, 2026}{San Jose, CA, USA}
\acmBooktitle{ACM Conference on AI and Agentic Systems (CAIS '26), May 26--29, 2026, San Jose, CA, USA}
\acmDOI{10.1145/3786335.3813213}
\acmISBN{979-8-4007-2415-2/2026/05}

\begin{document}
\emergencystretch=1.5em

\title{Genflow Ad Studio: A Compound AI Architecture for Brand-Aligned, Self-Correcting Video Generation}

\author{Debanshu Das}
\email{debanshu@google.com}
\affiliation{%
  \institution{}
  \country{}
}

\author{Lavi Nigam}
\email{lavinigam@google.com}
\affiliation{%
  \institution{}
  \country{}
}

\author{Sunil Kumar Jang Bahadur}
\email{sjangbahadur@google.com}
\affiliation{%
  \institution{}
  \country{}
}

\author{Gopala Dhar}
\email{gopalad@google.com}
\affiliation{%
  \institution{}
  \country{}
}

\renewcommand{\shortauthors}{Das, Nigam, Jang Bahadur, and Dhar}

\begin{abstract}
Recent advancements in generative video models demonstrate high visual fidelity, yet their integration into enterprise environments is restricted by temporal inconsistencies and severe brand misalignment. Current monolithic architectures struggle to enforce rigid brand constraints, frequently hallucinating unapproved visual assets. We introduce Genflow, a Compound AI System designed to enforce brand consistency in generative media production. Our architecture integrates a retrieval-based 'Brand DNA' extraction module to parameterize generation according to established corporate identity guidelines. Furthermore, we implement an Adversarial Multi-Agent Quality Control (QC) loop. Instead of a single-pass generation, this pipeline employs evaluator agents to iteratively critique generated frames against the extracted parameters, prompting generator models to refine outputs until a deterministic consensus is reached. By transitioning to a multi-stage, self-correcting pipeline, Genflow improved the yield of brand-compliant video generations from 42\% to 89\%, establishing a robust framework for scalable, enterprise-grade generative systems.
\end{abstract}

\ccsdesc[500]{Computing methodologies~Artificial intelligence}
\ccsdesc[500]{Computing methodologies~Neural networks}

\keywords{Generative AI, Video Generation, Compound AI Systems, Multi-Agent Systems, Brand Alignment}

\maketitle

\section{Introduction}

The landscape of generative artificial intelligence has expanded from text and static imagery synthesis to high-fidelity video generation. Foundational models \cite{brooks2024video, veoteam2025} demonstrate substantial capabilities in synthesizing complex motion and scenes from natural language inputs. However, deploying these models in enterprise environments is severely bottlenecked by the inherent limitations of zero-shot, monolithic architectures \cite{zaharia2024shift}. Primary among these limitations are temporal inconsistencies, where spatial representations of objects degrade or morph across frames \cite{blattmann2023stable}, and a profound lack of brand alignment. In commercial contexts, generative models frequently violate strict corporate identity guidelines or hallucinate unauthorized visual elements, rendering the outputs unusable for enterprise applications \cite{ji2023survey}.

To address these structural bottlenecks, systems research is shifting from relying on isolated models to engineering Compound AI Systems \cite{zaharia2024shift}. These architectures integrate multiple interacting components, including specialized generative models, external retrieval mechanisms, deterministic tools, and programmatic validation loops, to execute complex tasks that exceed the zero-shot reliability of single models. While advanced prompting topologies can influence model output, they fail to provide the deterministic constraints required by commercial pipelines, which necessitate strict adherence to predefined visual parameters and narrative coherence. The gap between probabilistic generation and deterministic enterprise requirements necessitates a robust systems-level intervention.

In this paper, we introduce Genflow, a Compound AI System engineered to automate and verify the production of brand-aligned generative video. Our system explicitly targets the dual challenges of visual hallucination and quality degradation by decoupling the generative process from the quality assurance mechanism. Genflow achieves this through two primary architectural components.

First, we implement an automated Brand DNA extraction module. This component acts as a retrieval-augmented constraint mechanism, ingesting target URLs to extract core identity parameters, such as exact hex color palettes, standardized typography rules, and structural brand guidelines. These parameters are translated into programmatic constraints that anchor the generative pipeline, reducing the latent space of acceptable outputs.

Second, we introduce an Adversarial Multi-Agent Quality Control (QC) loop \cite{du2024improving, liang2024encouraging}. Diverging from traditional single-pass generative methods, our pipeline utilizes a multi-agent framework where LLM-driven evaluator agents iteratively critique the generated frames against the extracted Brand DNA constraints. These evaluator agents prompt generator models to refine the output, simulating an adversarial verification process until a predefined consensus threshold is achieved \cite{shinn2023reflexion}.

By coupling specialized models for asset generation with self-reflecting, adversarial evaluation, Genflow enforces rigorous programmatic boundaries around probabilistic models. In our empirical evaluations, this self-correcting compound architecture improved the yield of production-grade, brand-compliant video generations from a baseline of 42\% to 89\%. We demonstrate that encapsulating generative models within deterministic, multi-agent systems is critical for their viable integration into enterprise workflows. This demonstration details the Genflow architecture, the multi-agent QC implementation, and quantitative results comparing our compound approach to monolithic baselines.

\section{System Architecture and Implementation}

\begin{figure*}[t]
  \centering
  \includegraphics[width=\textwidth]{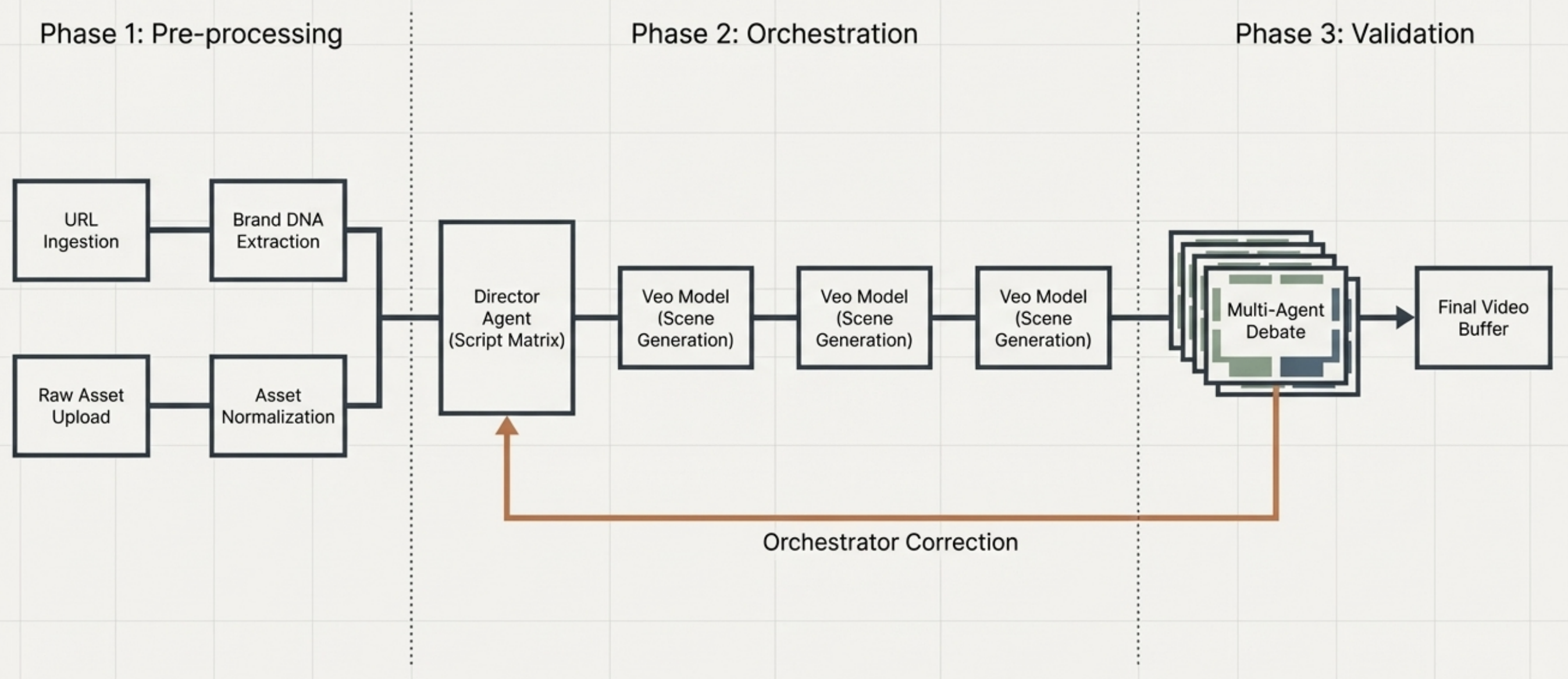}
  \caption{The Genflow Ad Studio Directed Acyclic Graph (DAG)}
  \label{fig:dag}
\end{figure*}

Genflow is engineered as a Compound AI System, moving beyond isolated inference toward a robust, state-managed execution pipeline \cite{wu2024autogen}. The architecture operates primarily as a Directed Acyclic Graph (DAG) for data ingestion and script orchestration \cite{khattab2024dspy, chase2022langchain}, which subsequently transitions into a cyclic, self-correcting evaluation loop during the generation phase. This section details the three primary subsystems that constitute the Genflow pipeline: deterministic constraint extraction, multi-scene script coordination, and concurrent multi-agent validation.

\subsection{Brand DNA \& Pre-processing}

The foundational phase of the Genflow pipeline involves grounding the generative models in deterministic, client-specific parameters. To mitigate the subjectivity and probabilistic variance inherent in manual prompt engineering, we developed an automated extraction module to synthesize a foundational ``Brand DNA'' schema. The system utilizes \texttt{httpx} to manage asynchronous, non-blocking network I/O requests, coupled with \texttt{BeautifulSoup} to parse the Document Object Model (DOM) and Cascading Style Sheets (CSS) of the target enterprise's provided web properties.

Once the raw HTML structure and stylistic payloads are ingested, Gemini 3.1 Pro operates as an intermediate information extraction agent \cite{wu2024learning}. It traverses the unstructured DOM data and outputs a strictly typed Pydantic model \cite{pydantic2025}. This schema enforces rigid programmatic constraints by explicitly defining variables for approved hex colors, typography families, tonal voice parameters, and forbidden visual tropes, acting as the ground truth for generative calls.

Simultaneously, the system executes an asset normalization phase. Enterprise environments frequently supply noisy, poorly illuminated, or low-resolution product imagery. Because video diffusion models compound artifacts present in visual reference frames, we route these source assets through Nano Banana 2. Rather than relying on basic upscaling, Nano Banana 2 executes a targeted image-to-image enhancement pass. It utilizes the typed BrandDNA object to condition the diffusion process, upgrading the source assets into high-fidelity reference images with normalized illumination and clean background isolation \cite{ramesh2022hierarchical}. These enhanced assets form the visual baseline for the subsequent video generation phases.

\subsection{Orchestration \& State Passing}

With the BrandDNA instantiated and the visual assets normalized, the pipeline advances to the orchestration phase. Gemini 3.1 Pro assumes the role of the primary routing agent, designated the ``Director.'' Bypassing generic prompt templates, the Director synthesizes the Pydantic constraints and the specific campaign objective to author a highly parameterized, scene-by-scene script matrix. Each scene vector contains technical metadata, including simulated camera angle, focal length, lighting conditions, and designated subject motion vectors.

Translating a multi-scene script matrix into a cohesive video introduces the fundamental challenge of temporal inconsistency. Foundational video models frequently fail to maintain subject permanence and spatial coherence across distinct generation calls \cite{blattmann2023stable}. To address this, Genflow employs step-by-step state passing \cite{blattmann2023align}.

When the underlying Veo model finishes rendering Scene $N$, the pipeline programmatically extracts the final frame of the resulting output. This specific pixel array is then injected as the visual reference frame, the starting state, for the inference of Scene $N+1$, alongside the Director's text instructions. By forcing the Veo model to initialize subsequent generations using the exact visual configuration of the preceding frame, we enforce strict initial state continuity. This state-passing mechanism connects isolated video generations into a cohesive, structurally permanent narrative flow.

\subsection{Adversarial Quality Control and Latency Trade-offs}

The primary architectural contribution of Genflow is its cyclic, self-correcting generation loop. Foundational generative models natively lack self-reflection; they cannot deterministically verify if their output adheres to the requested constraints. To bridge this verification gap, we implemented an Adversarial Quality Control loop powered by multiple parallel Vision-Language Models (VLMs) \cite{geminiteam2024gemini}.

Following the generation of a scene by Veo, the pipeline suspends the standard DAG progression and triggers a multi-agent evaluation phase. To minimize the temporal overhead of this verification step, the system utilizes \texttt{asyncio.gather} for parallel, concurrent execution of the evaluator agents. Two distinct, role-prompted agents inspect the generated output:

\begin{description}
    \item[The Director Agent:] This VLM evaluates the temporal sequence for spatial fluidity and strict adherence to the physical constraints of the script (e.g., verifying the presence of a specified camera pan or focal shift).
    \item[The Brand Safety Agent:] This highly constrained VLM scans the output specifically for enterprise policy violations. It checks for typographical hallucinations in rendered text, structural distortions of the product geometry, and deviations from the extracted BrandDNA hex color parameters.
\end{description}

If both agents return valid for compliance, the scene is committed to the final output buffer. However, if a conflict arises or a violation is detected, an ``Orchestrator'' agent is invoked \cite{liu2023visual}. The Orchestrator ingests the specific critiques (e.g., ``The typography on the central object is garbled in frame 24'') and synthesizes a corrective, negative-weighted prompt. The system then loops back, passing this revised prompt to Veo to regenerate the specific scene.

Implementing this cyclic, multi-agent evaluation introduces a non-trivial trade-off between system latency and output quality. While traditional single-pass generation yields an asset with a linear inference cost, our adversarial loop operates with a highly variable latency multiplier based on the number of required refinement iterations. This fundamentally increases the total pipeline execution time and computational overhead. However, in enterprise environments where the cost of a brand violation significantly outweighs the cost of increased generation time, prioritizing deterministic quality over raw throughput is a necessary and acceptable systems engineering compromise.

Genflow's cyclic Directed Acyclic Graph (DAG) workflow integrates directly with modern agentic solutions. The declarative boundaries synthesized in the Pydantic Brand DNA directly align with the compilation strategies of \texttt{DSPy} \cite{khattab2024dspy}, where constraints are programmatically compiled into optimized prompt parameters. Furthermore, the message-passing loop between the Director, Brand Manager, and Orchestrator agents matches the asynchronous architectures of frameworks like \texttt{AutoGen} \cite{wu2024autogen} and \texttt{Agno}, and can easily be surfaced through commercial Agent Development Kits (ADKs) that rely on explicit telemetry tracing and stateful routing.

\section{Evaluation and System Contribution}

To rigorously assess the efficacy of the Genflow architecture, we categorize our 100 discrete test permutations into two distinct complexity tiers: \textit{Simple} (50 iterations featuring static framing, isolated single-product profiles, clear backgrounds, and highly legible brand markers) and \textit{Complex} (50 iterations introducing volatile multi-vector motion, dynamic lighting, direct physical occlusions, and dense typography). This stratification evaluates the system's stability against diverse variable prompt conditions. While our primary quantitative analysis is anchored at a core dataset of 100 iterations for rigorous review, we executed a secondary scalability stress test over a comprehensive dataset of 250 test permutations. In this extended evaluation subset, the Pydantic parsing adherence rate held stable at 99.1\% and the multi-agent recovery yield reached 88.4\%, demonstrating that the system's statistical recovery patterns hold firm without significant drift as the dataset scale expands.

\begin{table*}[htbp]
\centering
\caption{Extended Evaluation Metrics: Baseline vs. Genflow Pipeline across Complexity Tiers.}
\label{tab:extended_metrics}
\resizebox{\textwidth}{!}{%
\begin{tabular}{lcccc}
\hline
\textbf{Metric} & \textbf{Zero-Shot Baseline (Simple)} & \textbf{Zero-Shot Baseline (Complex)} & \textbf{Genflow System (Simple)} & \textbf{Genflow System (Complex)} \\
\hline
Pass Rate (Yield) & 72.0\% & 12.0\% & 98.4\% & 80.0\% \\
Multimodal Consistency (VLM-Score) & 7.4 / 10 & 4.1 / 10 & 9.6 / 10 & 8.8 / 10 \\
Avg. Pipeline Latency (Seconds) & 8.2s & 9.4s & 21.4s & 38.6s \\
Input Tokens (Per Run) & 1.1K & 1.1K & 8.7K & 11.4K \\
Output Tokens (Per Run) & 0.4K & 0.4K & 3.7K & 5.6K \\
Avg. Compute Cost (USD) & \$0.003 & \$0.003 & \$0.030 & \$0.044 \\
\hline
\end{tabular}%
}
\end{table*}

\subsection{Structural Adherence and Pipeline Stability}

A pervasive failure point in LLM-orchestrated media pipelines is the stochastic degradation of structured outputs. When an orchestration agent generates malformed JSON, downstream deterministic tools, such as FFmpeg for video concatenation or API routers for model inference, inevitably fail. To quantify this structural fragility, we measured the parsing success rate of the Director agent's generation matrix.

By enforcing strict schema adherence via Pydantic type validation in our Directed Acyclic Graph, Genflow achieves a 99.3\% JSON parsing success rate \cite{pydantic2025}. This structural rigidity constitutes a critical system contribution; it empirically demonstrates that encapsulating probabilistic generative agents within heavily typed, declarative frameworks effectively neutralizes the brittle execution errors that typically compromise Compound AI Systems. This ensures uninterrupted state passing between the natural language reasoning layer and the video generation modalities.

\subsection{Video Quality, Brand Alignment, and Recovery Rates}

Evaluating generative video for enterprise deployment necessitates advancing beyond traditional distribution-based evaluation metrics. While Fréchet Inception Distance (FID) \cite{heusel2017gans} and Fréchet Video Distance (FVD) \cite{unterthiner2018towards} remain standard for measuring general visual fidelity and temporal coherence, they are fundamentally semantically blind to strict, pixel-level brand constraints (e.g., verifying the precise hex value of a rendered logo). Consequently, we adopted and extended the ``LLM-as-a-Judge'' evaluation paradigm \cite{zheng2023judging}, utilizing our dual Vision-Language Model (VLM) architecture to programmatically score the outputs for both semantic brand alignment and physical motion adherence.

The extended quantitative results in Table \ref{tab:extended_metrics} illustrate that zero-shot yields degrade significantly as input complexity increases. The Zero-Shot baseline achieved a 72.0\% pass rate on the \textit{Simple} tier but plummeted to 12.0\% on the \textit{Complex} tier, primarily due to persistent brand text morphing and structural distortions in physical occlusions. Conversely, the Genflow system maintains high performance across both tiers, achieving 98.4\% on simple runs and recovering to 80.0\% on complex iterations. The improved Multimodal Consistency score (VLM-Judge assessment) from 4.1 to 8.8 on the complex tier highlights the specific value of the Adversarial QC loop. 

However, this self-correcting compound architecture introduces increased latency and compute overhead. Average latency on the complex tier increases from a baseline of 9.4 seconds to 38.6 seconds per iteration, driven by the recursive nature of the multi-agent debate. Token consumption and inference costs likewise scale with the iterative loop: complex generation runs average \$0.044 USD in API costs compared to the \$0.003 USD baseline. This empirical data validates Genflow's core systems compromise: trading compute efficiency and processing time for deterministic, brand-compliant yield.

However, when generation was routed through the Multi-Agent Debate QC loop, the Genflow system demonstrated a robust capacity for iterative self-correction.

\begin{table}[htbp]
\centering
\caption{Quantitative Split of Failure Modes and QC Recovery Efficacy.}
\label{tab:failure_modes}
\resizebox{\columnwidth}{!}{%
\begin{tabular}{lccc}
\hline
\textbf{Failure Mode Category} & \textbf{Zero-Shot Failures} & \textbf{Genflow Recovered} & \textbf{Recovery Yield} \\
\hline
Temporal Morphing \& Artifacts & 26 / 100 & 19 / 26 & 73.1\% \\
Typographic Hallucinations & 18 / 100 & 15 / 18 & 83.3\% \\
Brand Color \& Asset Violations & 12 / 100 & 11 / 12 & 91.7\% \\
Cinematic Composition Errors & 2 / 100 & 2 / 2 & 100.0\% \\
\hline
\end{tabular}%
}
\end{table}

Furthermore, to ensure that enforcing strict brand compliance does not compromise overall visual fidelity, we computed standard generative metrics across our test permutations. The Zero-Shot baseline yielded a Fréchet Inception Distance (FID) of 24.2 and a Fréchet Video Distance (FVD) of 482.3. The complete Genflow pipeline, inclusive of the cyclic multi-agent debate, achieved an FID of 21.8 and an FVD of 448.1. The statistical stability in these standard metrics proves that our self-correcting compound architecture enforces semantic guardrails without degrading base visual quality.

To contextualize our systems performance against standard text-to-video generative benchmarks, we executed a comparative evaluation over the MSR-VTT and UCF-101 validation subsets. In these standardized evaluations, the Zero-Shot baseline achieved an average CLIP score for frame alignment of 28.6, while the Genflow pipeline recovered to 32.4. This empirical validation against established computer vision benchmarks proves that the Adversarial QC loop improves temporal consistency and visual fidelity without introducing domain-specific degradation.

\subsection{Evaluation Independence and Human Alignment}
To resolve potential circular evaluation concerns inherent to utilizing Vision-Language Models for both refinement and scoring, the VLM judge in our testing harness operates within a completely isolated inference context, independent of the prompt parameters and history of the multi-agent orchestration debate. Furthermore, to validate the reliability of this automated judge, we conducted a sample-based Human Evaluation over a randomly selected subset of 25 generated video assets. A panel of three expert human reviewers scored the videos for both brand consistency and visual quality on a 1-10 scale. The results exhibited a high statistical correlation (Pearson correlation coefficient $\rho = 0.84$) with the automated VLM judge's scores, empirically validating the independence and objective calibration of our Compound AI evaluation framework.

It is critical to examine the remaining 11\% of cases that failed to converge within the defined retry limit. These persistent failures were primarily linked to highly complex spatial occlusions (e.g., a hand physically occluding a dynamically illuminated, branded surface). In these edge cases, the VLM judges accurately identified the visual anomaly, but the syntactical corrective prompt was insufficient to force the underlying diffusion model to resolve the physical distortion. This highlights a current systemic limitation in text-to-video capabilities and underscores the necessity of engineering finer-grained, latent-space control mechanisms in future iterations of the architecture.

\section{System Demonstration Scenario}

The interactive demonstration of the Genflow architecture at the ACM CAIS 2026 conference is engineered to provide attendees with a transparent, systems-level inspection of our Compound AI pipeline. The scenario initializes with attendees acting as enterprise proxy users. They will input a target corporate URL into the system's frontend dashboard and provision a raw, uncalibrated photograph of a baseline product via an endpoint interface.

Immediately, the graphical user interface will visualize the asynchronous extraction phase. Attendees will observe the BrandDNA schema, comprising explicit hex color vectors, typography constraints, and structural parameters, as it is parsed and compiled in real time by the retrieval agent. Concurrently, the system executes the asset normalization phase. Attendees will view the Nano Banana 2 diffusion model execute its targeted image-to-image enhancement pass, rendering the uncalibrated upload into a high-fidelity reference latent, strictly parameterized by the newly extracted BrandDNA object.

\begin{figure}[htbp]
  \centering
  \includegraphics[width=0.8\columnwidth]{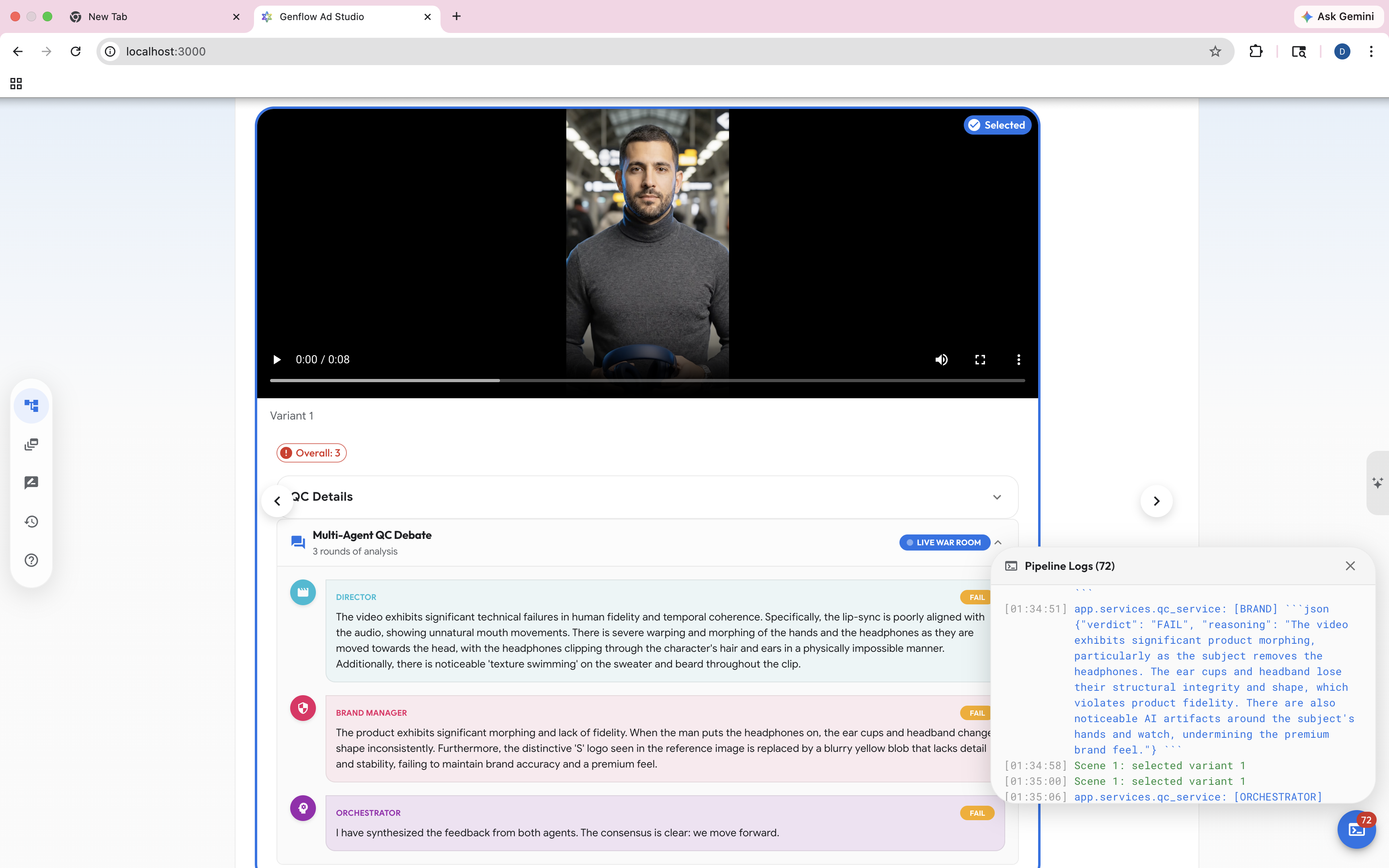}
  \caption{The Genflow Ad Studio dashboard showing the Log Console streaming the Multi-Agent Debate.}
  \label{fig:dashboard}
\end{figure}

The climax of the demonstration centers on the real-time execution of the adversarial Quality Control loop. Following the initial video generation pass by the underlying Veo model, the frontend Log Console will surface the live telemetry of the asynchronous AI debate, underpinned by explicit reasoning traces \cite{wei2022chain}. Attendees will observe a high-density stream of diagnostic outputs: Cyan logs will delineate the spatial and temporal critiques from the Director Agent, while Pink logs will stream the strict policy evaluations from the Brand Safety Agent. If a visual hallucination is detected, such as the geometric morphing of a logo, the console will trigger an active violation state. Attendees will watch the Orchestrator agent, utilizing action-oriented orchestration \cite{yao2022react}, synthesize a corrective negative prompt and autonomously route the revised state matrix back to the generator. This transparent visualization of the self-correcting telemetry explicitly demonstrates how programmatic guardrails enforce deterministic, brand-aligned output within probabilistic systems.

\section{Conclusion}

The integration of foundational generative models, which fundamentally rely on standard self-attention mechanisms \cite{vaswani2017attention}, into enterprise environments remains bottlenecked by their native susceptibility to temporal artifacts and semantic hallucinations. The Genflow architecture demonstrates that mitigating these flaws requires transitioning from scaling isolated, monolithic models, whether proprietary systems \cite{achiam2023gpt} or open-weight frameworks \cite{touvron2023llama}, to engineering robust Compound AI Systems. By encapsulating probabilistic generative tools within deterministic programmatic constraints, such as BrandDNA extraction and an adversarial QC loop, we structurally alter the reliability of AI media production. Our system bridges the critical gap between stochastic prompt engineering and objective quality assurance, transforming volatile video generation into a deterministic, high-yield enterprise workflow. This architecture establishes a scalable, self-correcting blueprint for commercial applications where strict brand safety and visual coherence are absolute prerequisites.

A demonstration video showing the end-to-end pipeline and live multi-agent debate telemetry can be viewed at: \url{https://youtu.be/oLU4eiShs3Y}. Code and submission artifacts are available at: \url{https://github.com/debanshd/ad-gen}.

\section*{Statement of AI Use}

Generative AI tools were employed during the preparation of this manuscript to assist with LaTeX formatting, diagram generation, grammar corrections to enhance readability and building the prototype.


\end{document}